\documentclass{PoS}
\usepackage[utf8]{inputenc}
\usepackage{amsmath}
\usepackage{amssymb}

\newcommand{\eq}[1]{\begin{eqnarray} #1 \end{eqnarray}}

\title{QCD equation of state at finite baryon density with fugacity expansion}

\ShortTitle{QCD EoS with fugacity expansion}

\author{\speaker{Volodymyr Vovchenko}\\
        Institut f\"ur Theoretische Physik,
Goethe Universit\"at Frankfurt, Frankfurt am Main, Germany\\
        Frankfurt Institute for Advanced Studies, Frankfurt am Main, Germany\\
        E-mail: \email{vovchenko@fias.uni-frankfurt.de}
        }
        
\author{Jan Steinheimer\\
        Frankfurt Institute for Advanced Studies, Frankfurt am Main, Germany
        }
        
\author{Owe Philipsen\\
         Institut f\"ur Theoretische Physik,
Goethe Universit\"at Frankfurt, Frankfurt am Main, Germany
        }
        
\author{Horst Stoecker\\
         Institut f\"ur Theoretische Physik,
Goethe Universit\"at Frankfurt, Frankfurt am Main, Germany\\
         Frankfurt Institute for Advanced Studies, Frankfurt am Main, Germany\\
         GSI Helmholtzzentrum f\"ur Schwerionenforschung GmbH, Darmstadt, Germany
        }

\abstract{
We explore the QCD equation of state at finite baryon density through an expansion in baryon number fugacity, making use of the recent lattice data on the four leading Fourier coefficients of the expansion.
A state-of-the-art description of the lattice data at both imaginary $\mu_B$ and $\mu_B = 0$ is provided by the cluster expansion model~(CEM).
The CEM pressure function has three temperature dependent input parameters, which we fix by parameterizing the available lattice QCD data.
This results in a crossover model of the QCD equation of state at finite baryon density, which can be used in fluid dynamical simulations of heavy-ion collisions.
}

\FullConference{Corfu Summer Institute 2018 "School and Workshops on Elementary Particle Physics and Gravity"\\
		(CORFU2018)\\
		31 August - 28 September, 2018\\
		Corfu, Greece}

\begin{document}

\section{Introduction}

The equation of state of hot QCD matter is available for $\mu_B = 0$ from first-principle lattice QCD simulations~\cite{Borsanyi:2013bia,Bazavov:2014pvz}.
Finite $\mu_B$ simulations, on the other hand, are hindered by the sign problem. 
All the available lattice methods here are indirect and restricted to small values of $\mu_B/T$. 
These methods include the analytic continuation from imaginary chemical potential~\cite{deForcrand:2002hgr,DElia:2002tig}, the Taylor expansion around $\mu_B = 0$~\cite{Allton:2002zi,Gavai:2003mf}, the reweighing technique~\cite{Barbour:1997ej,Fodor:2001au}, and the canonical approach~\cite{Hasenfratz:1991ax,Nagata:2010xi}.
At larger baryon density the equation of state is usually described by effective models.

Recent advances in lattice QCD simulations at zero and imaginary chemical potential include the evaluation of higher-order baryon number susceptibilities at $\mu_B = 0$~\cite{DElia:2016jqh,Bazavov:2017dus,Borsanyi:2018grb} and the Fourier coefficients of net baryon density at imaginary $\mu_B$~\cite{Vovchenko:2017xad}, which put stringent constraints on effective models for the QCD equation of state at finite baryon density.
Here we construct a model for the QCD equation of state at finite baryon density which incorporates all of the  above-mentioned constraints.
Our considerations are based on the recently developed cluster expansion model~(CEM) formalism~\cite{Vovchenko:2017gkg,Vovchenko:2018zgt}. It uses the cluster expansion in baryonic fugacity,
\eq{
\frac{p(T,\mu_B)}{T^4} = \frac{1}{2} \sum_{k=-\infty}^{\infty} \, p_{|k|}(T) \, e^{k \, \mu_B/T},
}
which represents the pressure as a Laurent series in $\lambda_B \equiv e^{\mu_B/T}$.
This expansion incorporates two important QCD symmetries: the CP-symmetry~($\mu_B \to -\mu_B$) and the Roberge-Weiss periodicity~($\mu_B \to \mu_B + i \, 2 \pi T$) of the partition function.
The net baryon density reads
\eq{\label{eq:rhoBdef}
\frac{\rho_B(T,\mu_B)}{T^3} & = & \frac{1}{2} \sum_{k=1}^{\infty} \, b_{k}(T) \, \left[e^{k \, \mu_B/T} - e^{-k \, \mu_B/T}\right] \\
& = & \sum_{k=1}^{\infty} b_k(T) \, \sinh\left( \frac{k \mu_B}{T} \right), \qquad b_k \equiv k \, p_k.
}

At purely imaginary baryochemical potential the net baryon density has the form of a trigonometric Fourier series expansion
\eq{
\left. \frac{\rho_B(T,\mu_B)}{T^3} \right|_{\mu_B = i \, \theta_B \, T} & = & i \, \sum_{k=0}^{\infty} b_k(T) \, \sin\left( \frac{k \mu_B}{T} \right).
}
The expansion coefficients $b_k$ become Fourier expansion coefficients which can be extracted through the standard Fourier analysis:
\eq{\label{eq:FourierDef}
b_k(T) = \frac{2}{\pi} \int_0^{\pi} {\rm Im} \left[ \frac{\rho_B(T, i \theta_B \, T)}{T^3} \right] \, \sin(k \, \theta_B) \, d \theta_B~.
}
These Fourier coefficients can be evaluated in lattice QCD simulations at imaginary chemical potential~\cite{Lombardo:2006yc,Bornyakov:2016wld}. 
The four leading coefficients were evaluated at the physical point in Ref.~\cite{Vovchenko:2017xad} in the temperature range $135 < T < 230$~MeV.

\section{Cluster expansion model}

\subsection{Formulation}

The CEM~\cite{Vovchenko:2017gkg,Vovchenko:2018zgt} is a model for the QCD equation of state at finite baryon density which makes use of the expansion in fugacities~\eqref{eq:rhoBdef}.
It is based on an empirical observation of temperature independent Fourier coefficient ratios
\eq{\label{eq:alphak}
\alpha_k = \frac{[b_1(T)]^{k-2}}{[b_2(T)]^{k-1}} \, b_k(T), \qquad k = 3,4,\ldots
}
as observed in the lattice data of Ref.~\cite{Vovchenko:2017xad} for $k = 3,4$.
The available lattice data are consistent with a temperature independent $\alpha_{3,4}$ values, as evaluated in the Stefan-Boltzmann limit of massless quarks:
\eq{\label{eq:bkSB}
b_k^{\rm SB} & = & \frac{(-1)^{k-1}}{k} \, \frac{4 \, [3 + 4 \, (\pi k)^2]}{27 \, (\pi k)^2}, \\
\label{eq:alphakSB}
\alpha_k^{\rm SB} & = & 8^{k-1} \, \frac{(3+4\pi^2)^{k-2}}{(3+16\pi^2)^{k-1}} \, \frac{3+4\pi^2 k^2}{k^3}, \qquad k = 3,4,\ldots
}
The comparison of the lattice data for $\alpha_3$ and $\alpha_4$ with Eq.~\eqref{eq:alphakSB} is shown in Fig.~\ref{fig:alphak}.

The CEM assumes $\alpha_k = \alpha_k^{\rm SB}$ for all $k \geq 3$.
This implies that all higher-order Fourier coefficients are given in the CEM in terms of the two leading ones:
\eq{\label{eq:bkCEM}
b_k (T) = \alpha_k^{\rm SB} \, \frac{[b_2(T)]^{k-1}}{[b_1(T)]^{k-2}}, \qquad k = 3,4,\ldots
}
The expansion in Eq.~(\ref{eq:rhoBdef}) can be analytically summed in the CEM~(\ref{eq:bkCEM}).
The result is
\begin{equation}
\label{eq:analyt}
\frac{\rho_B(T,\mu_B)}{T^3} = 
-\frac{2}{27 \pi^2} \, \frac{\hat{b}_1^2}{\hat{b}_2} \, \left\{ 4 \pi^2  \, \left[ \textrm{Li}_1\left(x_+\right) - \textrm{Li}_1\left(x_-\right) \right] + 3 \, \left[ \textrm{Li}_3\left(x_+ \right) - \textrm{Li}_3\left(x_-\right) \right] \right\}.
\end{equation}
Here $\hat{b}_{k} = \displaystyle \frac{b_{k}(T)}{b_{k}^{\rm SB}}$, $x_{\pm} = \displaystyle - \frac{\hat{b}_2}{\hat{b}_1} \, e^{\pm \mu_B/T}$, and $\textrm{Li}_s(z) = \displaystyle \sum_{k=1}^{\infty} \frac{z^k}{k^s}$ is the polylogarithm.

\subsection{Comparison with the rational function model}

In general, a determination of the higher-order Fourier coefficients from the two leading ones is not unique.
Functional forms different from \eqref{eq:bkCEM} can be considered.
This point has been raised in Ref.~\cite{Almasi:2018lok}, where a rational function model~(RFM) has been introduced.
The Fourier coefficients in the RFM are given as follows:
\eq{\label{eq:bkrfm}
\hat{b}_k^{\rm RFM} (T) = \frac{c(T)}{1+k/k_0(T)}, \qquad k = 3,4,\ldots
}
with
\eq{
k_0(T) & = & [\hat{b}_1/\hat{b}_2 - 1]^{-1} - 1, \\
c(T) & = & \hat{b}_1(T) \, [1 + 1/k_0(T)].
}

\begin{figure}[t]
  \centering
  \includegraphics[width=.65\textwidth]{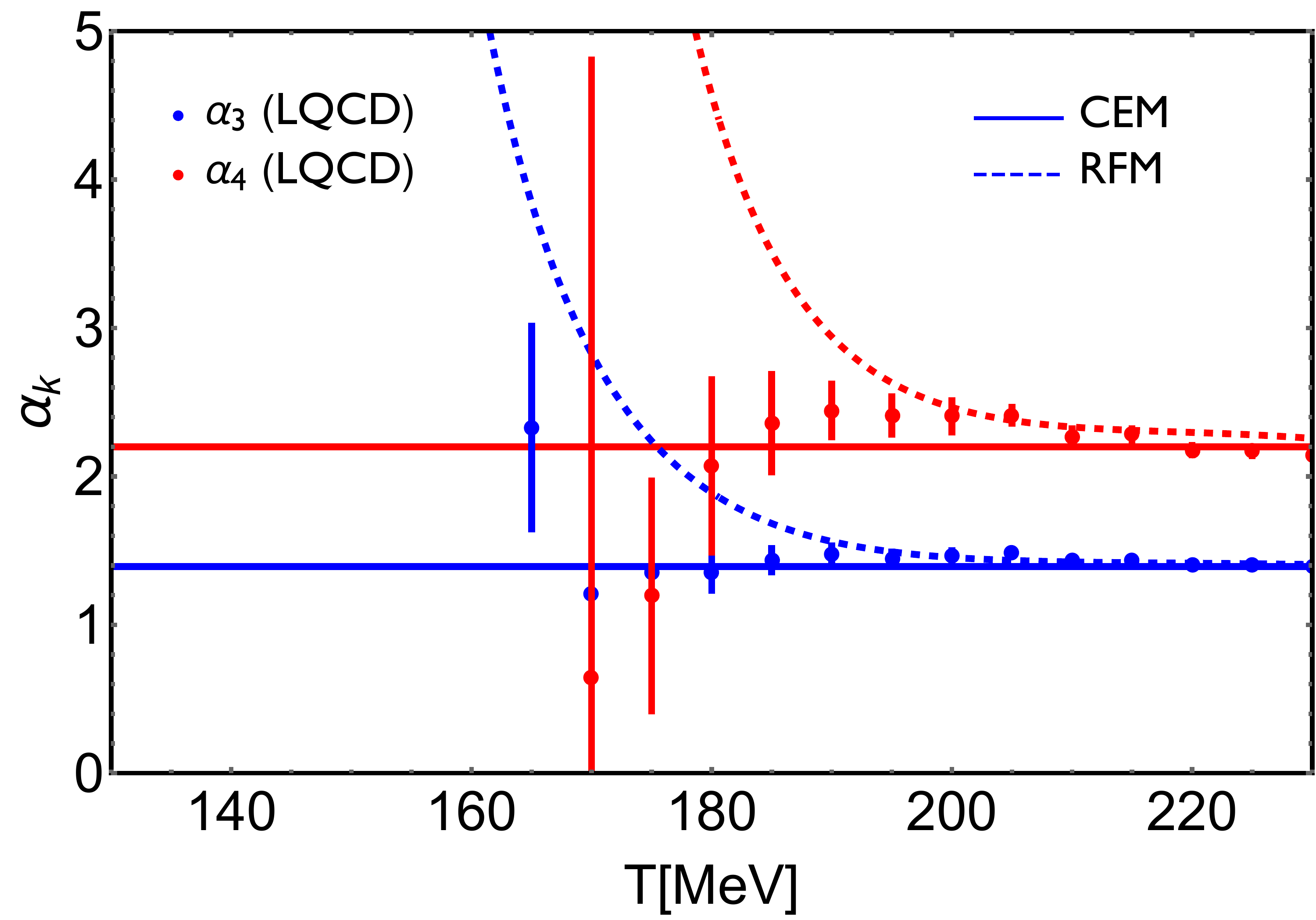}
  \caption{
   Temperature dependence of the Fourier coefficient ratios $\alpha_3$~(blue lines) and $\alpha_4$~(red lines) calculated within the CEM~(solid lines) and the rational function model of Ref.~\cite{Almasi:2018lok}~(dashed lines).
   Lattice data of the Wuppertal-Budapest collaboration~\cite{Vovchenko:2017xad} are depicted by the symbols with error bars.
  }
  \label{fig:alphak}
\end{figure}

The coefficients in the RFM exhibit a different asymptotic behavior in the temperature range considered: a power-law damping at large $k$ instead of an exponential damping in the CEM.
The RFM gives a similarly good description of the lattice data for the coefficients $b_3(T)$ and $b_4(T)$ as the CEM when these are viewed on the linear scale~(see Ref.~\cite{Almasi:2018lok}).
However, significant differences can be seen by considering the $\alpha_3$ and $\alpha_4$ ratios~[Eq.~\eqref{eq:alphak}], which are plotted for the CEM and the RFM in Fig.~\ref{fig:alphak}.
While the two models describe similarly well the lattice data at $T \gtrsim 200$~MeV, the RFM shows increased values of $\alpha_3$ and $\alpha_4$ at $T \lesssim 190$~MeV, which is not supported by the available data.

The subtle differences in the higher-order Fourier coefficients between the two models may show up in other observables, in particular those sensitive to the equation of state at finite baryon density.
Particularly interesting ones are the baryon number susceptibilities at $\mu_B = 0$, which can formally be given in terms of a series of the Fourier coefficients:
\eq{\label{eq:chis}
\chi_{2n}^{B} (T) \equiv \left. \frac{\partial^{2n-1} (\rho_B/T^3)}{\partial (\mu_B/T)^{2n-1}} \right|_{\mu_B = 0} = \sum_{k=1}^{\infty} \, k^{2n-1} \, b_k(T).
}
The sum~\eqref{eq:chis} can be carried out analytically in both the CEM and the RFM~(see Refs.~\cite{Vovchenko:2018zgt} and \cite{Almasi:2018lok}, respectively, for the details).

\begin{figure}[t]
  \centering
  \includegraphics[width=.49\textwidth]{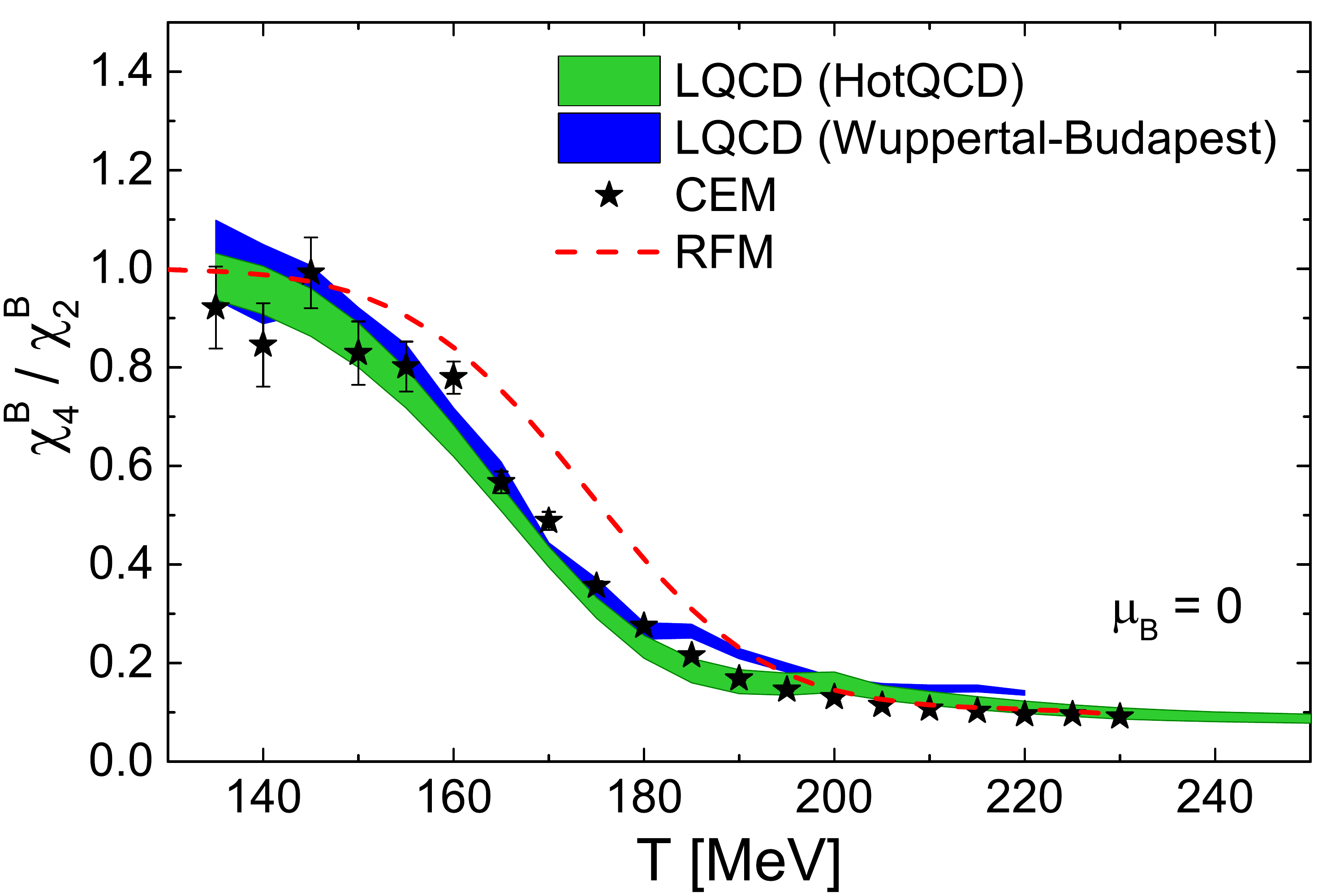}
  \includegraphics[width=.49\textwidth]{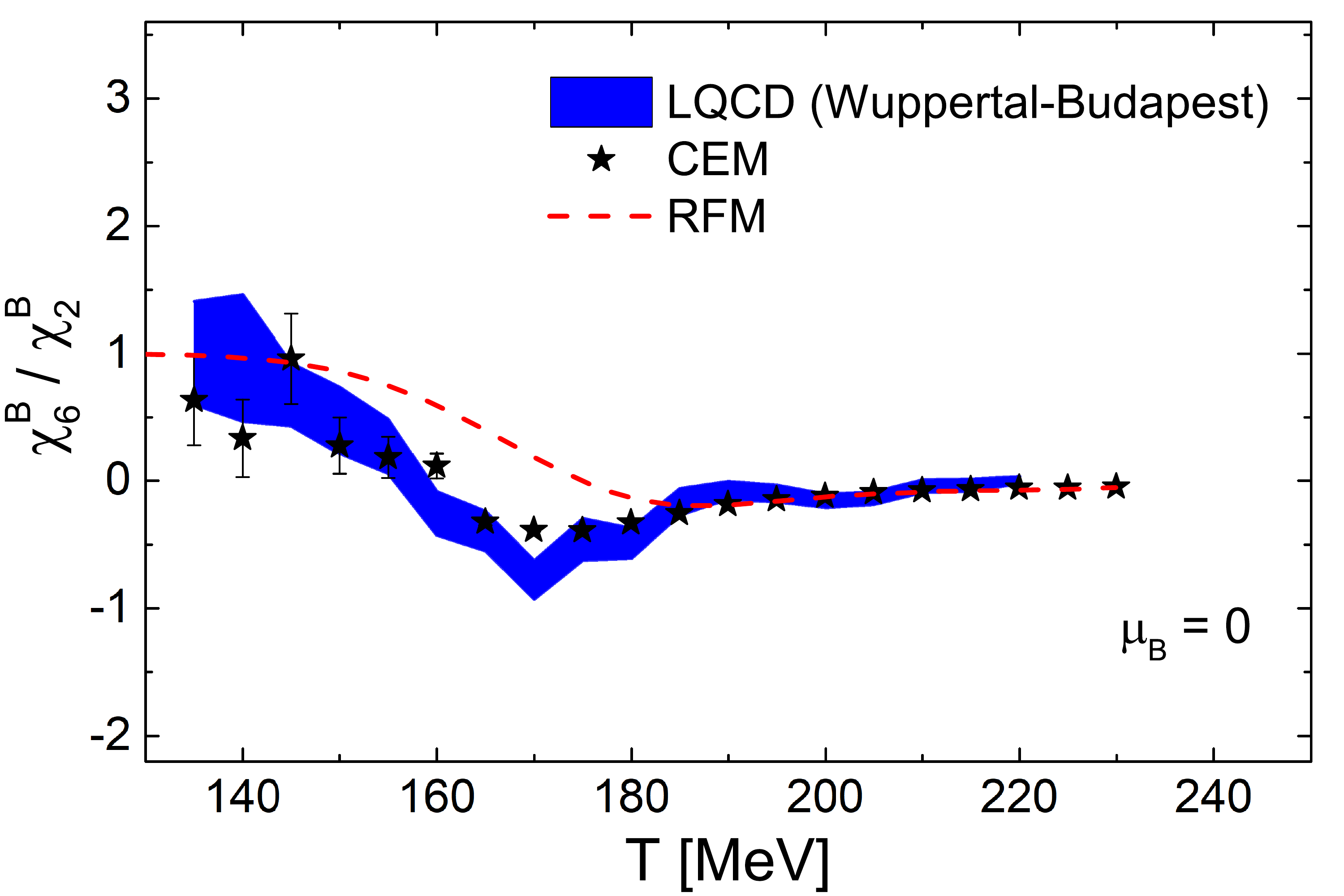}
  \caption{
   Temperature dependence of the baryon number susceptibility ratios $\chi_4^B/\chi_2^B$~(left panel) and $\chi_6^B/\chi_2^B$~(right panel), as calculated within the CEM~(black stars) and the rational function model of Ref.~\cite{Almasi:2018lok}~(dashed red lines). Lattice data of the Wuppertal-Budapest~\cite{Borsanyi:2018grb} and HotQCD collaborations~\cite{Bazavov:2017dus} are depicted by the blue and green bands, respectively.
  }
  \label{fig:chi2chi4chi6}
\end{figure}

Figure~\ref{fig:chi2chi4chi6} depicts the temperature dependence of the susceptibility ratios $\chi_4^B/\chi_2^B$ and $\chi_6^B/\chi_2^B$, as calculated within the CEM~(red stars) and the RFM~(dashed red lines) along with the lattice data of the Wuppertal-Budapest~\cite{Borsanyi:2018grb} and HotQCD collaborations~\cite{Bazavov:2017dus}.
The two models yield similar results in the low and high temperature limits while a notable difference is seen in the range $150 \lesssim T \lesssim 180$~MeV.
This difference can already be distinguished within the precision of the available lattice data, which appears to favor the CEM among the two models considered.
The CEM also describes fairly well the available lattice estimates for $\chi_8^B$~\cite{Borsanyi:2018grb}~(see Ref.~\cite{Vovchenko:2018zgt} for the agreement level).

While a determination of the higher-order Fourier coefficients from the lower order ones is not unique, we conclude that the CEM appears to be the best available tool for constructing a model of the QCD equation of state which is consistent with both the $\mu_B = 0$ and imaginary $\mu_B$ lattice data for physical quark masses.
We, therefore, proceed with constructing the full pressure function $p(T,\mu_B)$ within the CEM framework.

\section{The CEM equation of state}

The CEM pressure is obtained by integrating the net baryon density~\eqref{eq:analyt} over $\mu_B$:
\eq{\label{eq:CEMp}
\frac{p(T,\mu_B)}{T^4} & = & \frac{p_0(T)}{2} -\frac{2}{27 \pi^2} \, \frac{\hat{b}_1^2}{\hat{b}_2} \, \left\{ 4 \pi^2  \, \left[ \textrm{Li}_2\left(x_+\right) - \textrm{Li}_2\left(x_-\right) \right] + 3 \, \left[ \textrm{Li}_4\left(x_+ \right) - \textrm{Li}_4\left(x_-\right) \right] \right\}, \nonumber \\
& = & \frac{p_0(T)}{2} + \frac{\Delta p (T, \mu_B)}{T^4}.
}
The CEM input parameters are the temperature-dependent coefficients $p_0(T)$, $b_1(T)$, and $b_2(T)$.
At high temperatures~($T \gtrsim 130$~MeV) these can be extracted from the available lattice data.
At low temperatures they can be matched to the equation of state of the hadron resonance gas~(HRG) model.
This matching is achieved through a smooth switching function~\cite{Albright:2014gva}.

The first Fourier coefficient, $b_1(T)$, is parameterized as follows:
\eq{
b_1(T) & = & \left[1 - S(T) \right] \, b_1^{\rm hrg}(T) + S(T) \, b_1^{\rm lat} (T), \\
S(T) & = & \exp\left[\left(-\frac{T}{T_0}\right)^{-r}\right].
}
Here $b_1^{\rm hrg}$ is the partial pressure of baryons and antibaryons at $\mu_B = 0$ in the ideal hadron resonance gas~(HRG) model. 
We evaluate it using the HRG model of Ref.~\cite{Vovchenko:2017xad}.
$b_1^{\rm lat}$ corresponds to the lattice data of the Wuppertal-Budapest collaboration~\cite{Vovchenko:2017xad} at sufficiently high temperatures. We parametrize $b_1^{\rm lat}(T)$ as
\eq{
b_1^{\rm lat} (T) = \frac{b_1^{\rm sb} + a_n/t_b + b_n/t_b^2}{1 + a_d/t_b + b_d / t_b^2}, \qquad t_b = T/T_0~,
}
with the following parameter values:
\eq{
a_n = -0.940, \quad b_n = 0.345, \quad a_d = -1.336, \quad b_d = 0.502.
}
The values of the parameters $T_0$ and $r$ are listed in the first row of Table~\ref{tab:params}.

The second Fourier coefficient, $b_2(T)$, is represented in a form
\eq{
b_2(T) = -[b_1(T)]^2 \, T^3 \, b(T).
}
The parameter $b(T)$ can be interpreted as a baryonic eigenvolume parameter in a hadron gas with repulsive baryonic interactions~\cite{Vovchenko:2017xad}.
We observe that the lattice data for $b(T)$~\cite{Vovchenko:2017xad} can be well described at high temperatures by a power-law dependence, $b(T) \sim T^{-3}$.
At low temperatures we assume that the matter corresponds to a HRG with residual repulsive baryonic interactions, which are characterized by the eigenvolume parameter value of $b^{\rm hrg} = 1$~fm$^3$.
The complete temperature dependence of $b$ is, therefore, parametrized as
\eq{
b(T) = \left[1 - S(T) \right] \cdot b^{\rm hrg} + S(T) \left[-\frac{b_2^{\rm sb}}{(b_1^{\rm sb})^2 \, T^3} \right].
}
The values of the switching function parameters $T_0$ and $r$ for $b(T)$ are listed in the second row of Table~\ref{tab:params}.

\begin{table}[t]
\begin{center}
\begin{tabular}{|l|c|c|c|}
\hline
        & $b_1(T)$ & $b(T)$  & $p_0(T)$  \\ \hline
$T_0$ [MeV] & 170   & 175.5 & 100 \\
$r$         & 7.95       & 8     & 7   \\ \hline
\end{tabular}
\end{center}
\label{tab:params}
\caption{
Values of the switching function parameters $T_0$ and $r$ for the parameterizations of $b_1(T)$, $b(T)$, and $p_0(T)$.
}
\end{table}

In order to parameterize $p_0(T)$ we use the lattice QCD data for the pressure at zero chemical potential, as well the already obtained parameterizations for $b_{1,2}(T)$. 
At small temperatures $p_0(T)$ is matched with the doubled partial pressure of mesons in an ideal HRG model.
A complete parameterization is thus given by the following:
\eq{
p_0(T) = \left[1 - S(T) \right] p_0^{\rm hrg} + S(T) \, 2 \left[ \frac{p^{\rm lat}(T,\mu_B=0) - \Delta p (T, \mu_B = 0)}{T^4} \right].
}
Here $p^{\rm lat}(T,\mu_B=0)$ is evaluated using the $p/T^4$ parametrization of the HotQCD collaboration from Ref.~\cite{Bazavov:2014pvz} whereas $\Delta p (T, \mu_B = 0)$ depends on $b_1(T)$ and $b_2(T)$ only, both of which are already parametrized.
The values of the switching function parameters $T_0$ and $r$ for $b(T)$ are listed in the third row of Table~\ref{tab:params}.

Note that within the parametrization considered, $p_0$, $b_1$, and $b_2$ all smoothly approach the Stefan-Boltzmann limit of massless quarks and gluons in the high-temperature limit.

\begin{figure}[t]
  \centering
  \includegraphics[width=.49\textwidth]{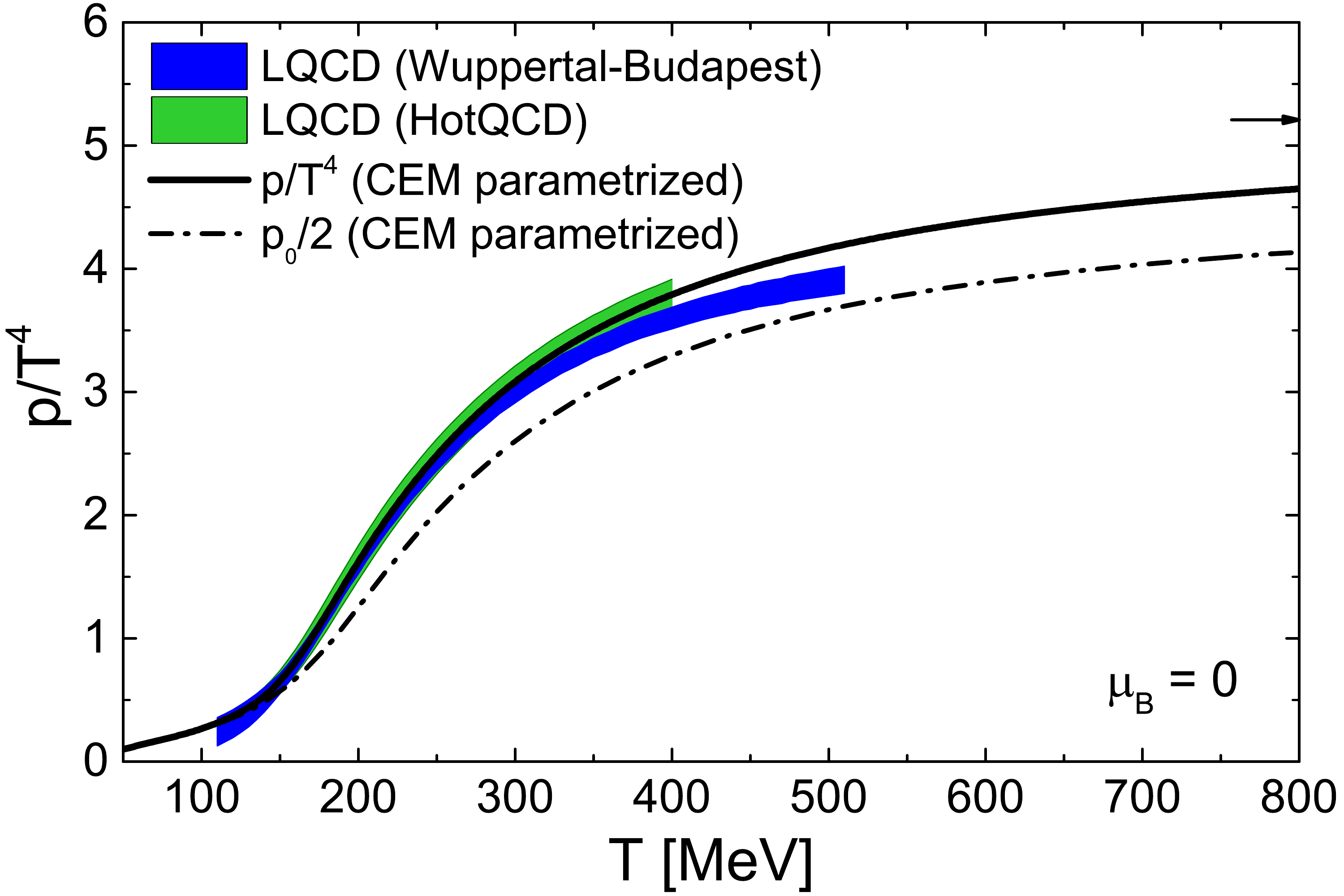}
  \includegraphics[width=.49\textwidth]{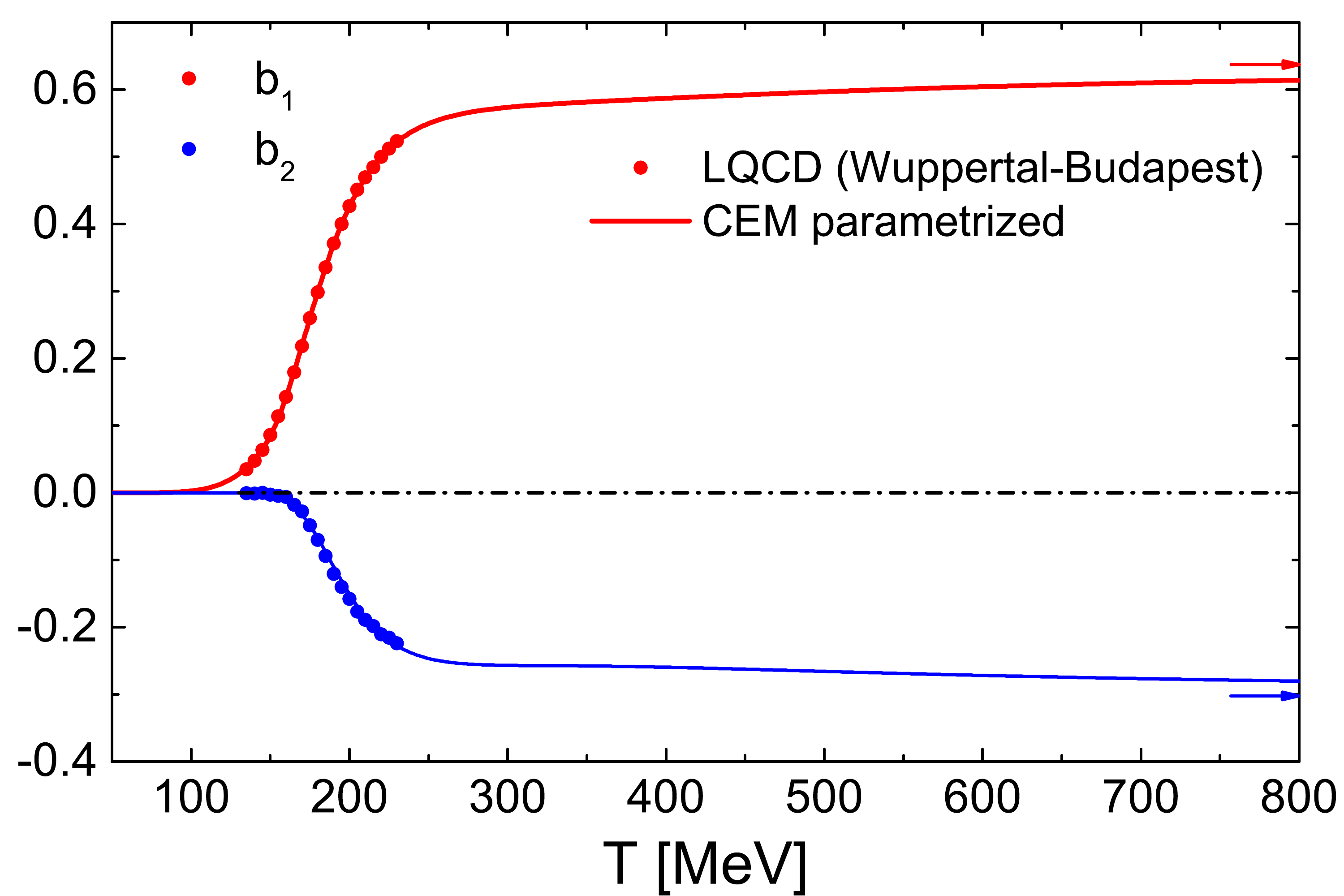}
  \caption{Temperature dependence of the scaled pressure and the pressure contribution $p_0/2$ of the 0th Fourier harmonic~(left panel), and the Fourier coefficients $b_1$ and $b_2$~(right panel), as obtained in the parametrized CEM. The lattice data of the Wuppertal-Budapest~\cite{Borsanyi:2013bia,Vovchenko:2017xad} and the HotQCD collaborations~\cite{Bazavov:2014pvz} are depicted where available.
  }
  \label{fig:paramsTdep}
\end{figure}

The temperature dependence of the resulting pressure $p/T^4$ at $\mu_B = 0$, the pressure contribution $p_0/2$ of the 0th Fourier harmonic, and the Fourier coefficients $b_1$ and $b_2$ is depicted in Fig.~\ref{fig:paramsTdep} for a broad temperature range $50 < T < 800$~MeV.
It is seen that $p_0/2$ corresponds to the bulk of the total pressure at $\mu_B = 0$ across all temperatures. 
Note that $p_0/2$ can be associated with the partial pressure of mesons at sufficiently small temperatures, where a HRG picture can be applied.

\begin{figure}[t]
  \centering
  \includegraphics[width=.90\textwidth]{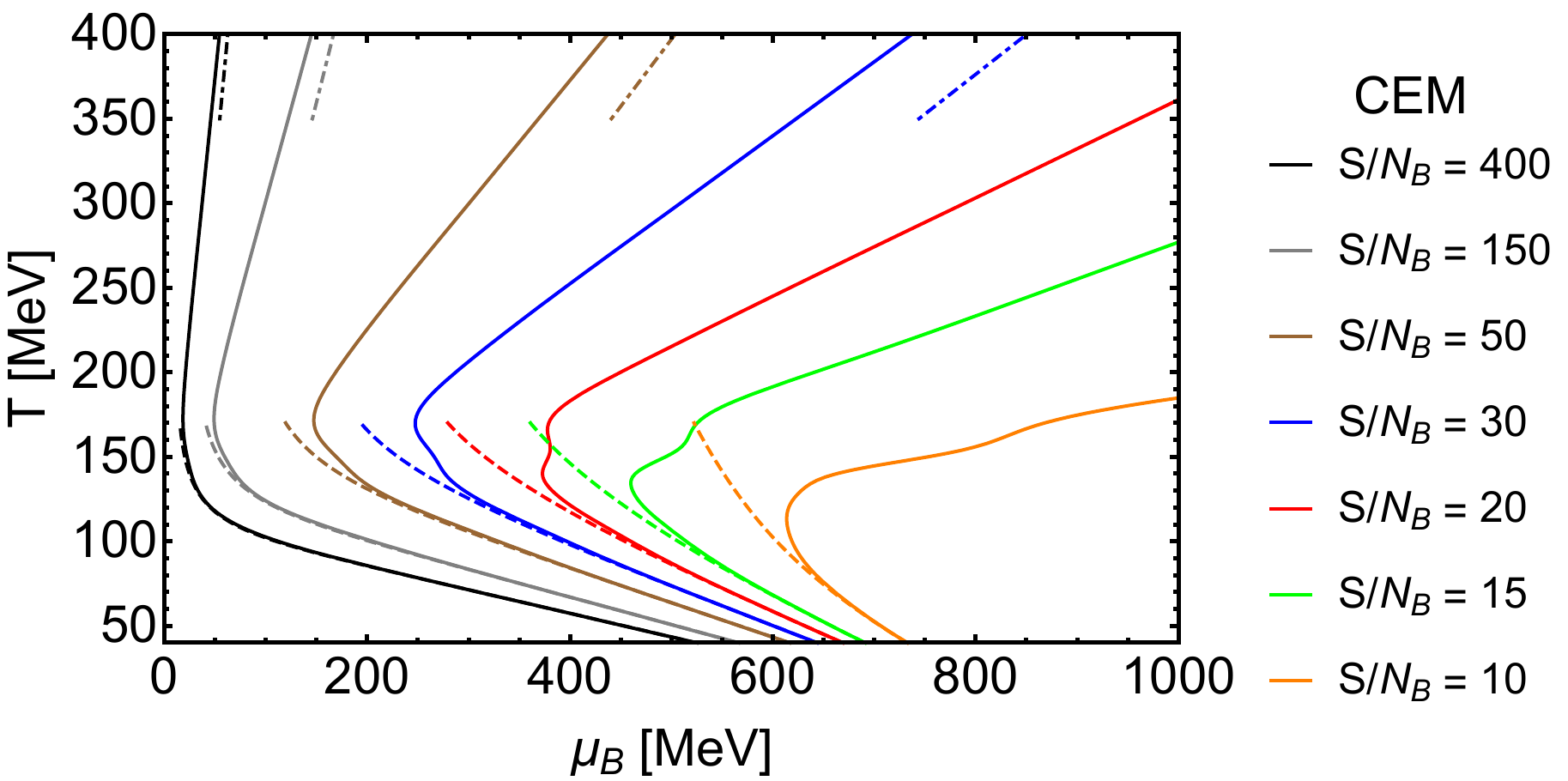}
  \caption{
   Isentropic trajectories on the $T$-$\mu_B$ plane, evaluated using the parameterized CEM equation of state for different entropy-per-baryon values~(solid lines). 
   The isentropes of the ideal HRG model are depicted by the dashed lines.
   The isentropes in the Stefan-Boltzmann limit of massless quarks and gluons are depicted by the dash-dotted lines.
  }
  \label{fig:SoverA}
\end{figure}

Different isentropic trajectories in the $\mu_B$-$T$ plane are depicted in Fig.~\ref{fig:SoverA} to illustrate the behavior of the CEM equation of state at finite baryon densities for conditions which can be expected to be created in relativistic heavy-ion collisions at various energies.
Seven entropy-per-baryon values are considered, $S/N_B = 10$, 15, 20, 30, 50, 150, and 400, which cover collision energies ranging from low SPS to top RHIC.
The overall behavior of the isentropes is reasonable, smoothly interpolating between the isentropes of an ideal HRG model~(dashed lines in Fig.~\ref{fig:SoverA}) and those of an ideal massless gas of quarks and gluons~(dot-dashed lines in Fig.~\ref{fig:SoverA}).

The CEM equation of state yields a broad crossover transition across the whole $\mu_B$-$T$ plane. 
The model, therefore, corresponds to the no-critical-point scenario.
It should be noted that the model is not suited for the cold and dense region of nuclear matter at low temperatures, as noted in Ref.~\cite{Vovchenko:2017gkg}.
A considerably more involved approach might be necessary to incorporate contraints from both the nuclear matter properties and lattice QCD data at $\mu_B = 0$ simultaneously~\cite{Motornenko:2019arp}.

\section{Summary}

The QCD equation of state at finite baryon density, based on the parameterized form of the cluster expansion model, has been presented.
This provides a first example of an equation-of-state model, which is originally formulated at purely imaginary baryochemical potential, and then analytically continued to real values of $\mu_B$.
A particular feature of the CEM equation of state is that it is able to incorporate constraints from both the lattice simulations at $\mu_B = 0$~(net baryon number susceptibilities) and imaginary $\mu_B$~(Fourier coefficients of net baryon density) simultaneously.
The model provides a reasonable behavior of the isentropes in the $\mu_B$-$T$ plane and can be used in fluid dynamical simulations of heavy-ion collisions from the lowest SPS energies up to the highest LHC energies.
The CEM equation of state presented here is available in a tabulated format~\cite{CEMtable}.

The present CEM equation of state exhibits a regular behavior at real values of $\mu_B$ and thus corresponds to the no-critical-point scenario for the QCD phase diagram, exhibiting a crossover-type transition all over the $\mu_B$-$T$ plane.
A generalization of the present model to incorporate a critical point at a certain location at finite baryon density is a possible future endeavor.
It should also be noted that the CEM is currently restricted to the baryochemical potential only.
A generalization of the approach to include the strangeness and electric charge chemical potentials is another possible future avenue to explore.

\section*{Acknowledgments}
H.St. acknowledges
the support through the Judah M. Eisenberg Laureatus Chair at Goethe University, and the Walter Greiner
Gesellschaft, Frankfurt.


\begin{thebibliography}{99}

\bibitem{Borsanyi:2013bia}
  S.~Borsanyi, Z.~Fodor, C.~Hoelbling, S.~D.~Katz, S.~Krieg and K.~K.~Szabo,
  Phys.\ Lett.\ B {\bf 730} (2014) 99
  [arXiv:1309.5258 [hep-lat]].


\bibitem{Bazavov:2014pvz}
  A.~Bazavov {\it et al.} [HotQCD Collaboration],
  Phys.\ Rev.\ D {\bf 90} (2014) 094503
  [arXiv:1407.6387 [hep-lat]].


\bibitem{deForcrand:2002hgr}
  P.~de Forcrand and O.~Philipsen,
  Nucl.\ Phys.\ B {\bf 642} (2002) 290
  [hep-lat/0205016].


\bibitem{DElia:2002tig}
  M.~D'Elia and M.~P.~Lombardo,
  Phys.\ Rev.\ D {\bf 67} (2003) 014505
  [hep-lat/0209146].


\bibitem{Allton:2002zi}
  C.~R.~Allton, S.~Ejiri, S.~J.~Hands, O.~Kaczmarek, F.~Karsch, E.~Laermann, C.~Schmidt and L.~Scorzato,
  Phys.\ Rev.\ D {\bf 66} (2002) 074507
  [hep-lat/0204010].


\bibitem{Gavai:2003mf}
  R.~V.~Gavai and S.~Gupta,
  Phys.\ Rev.\ D {\bf 68} (2003) 034506
  [hep-lat/0303013].


\bibitem{Barbour:1997ej}
  I.~M.~Barbour, S.~E.~Morrison, E.~G.~Klepfish, J.~B.~Kogut and M.~P.~Lombardo,
  Nucl.\ Phys.\ Proc.\ Suppl.\  {\bf 60A} (1998) 220
  [hep-lat/9705042].


\bibitem{Fodor:2001au}
  Z.~Fodor and S.~D.~Katz,
  Phys.\ Lett.\ B {\bf 534} (2002) 87
  [hep-lat/0104001].


\bibitem{Hasenfratz:1991ax}
  A.~Hasenfratz and D.~Toussaint,
  Nucl.\ Phys.\ B {\bf 371} (1992) 539.


\bibitem{Nagata:2010xi}
  K.~Nagata and A.~Nakamura,
  Phys.\ Rev.\ D {\bf 82} (2010) 094027
  [arXiv:1009.2149 [hep-lat]].


\bibitem{DElia:2016jqh}
  M.~D'Elia, G.~Gagliardi and F.~Sanfilippo,
  Phys.\ Rev.\ D {\bf 95} (2017) 094503
  [arXiv:1611.08285 [hep-lat]].


\bibitem{Bazavov:2017dus}
  A.~Bazavov {\it et al.},
  Phys.\ Rev.\ D {\bf 95} (2017) 054504
  [arXiv:1701.04325 [hep-lat]].


\bibitem{Borsanyi:2018grb}
  S.~Borsanyi, Z.~Fodor, J.~N.~Guenther, S.~K.~Katz, K.~K.~Szabo, A.~Pasztor, I.~Portillo and C.~Ratti,
  JHEP {\bf 1810} (2018) 205
  [arXiv:1805.04445 [hep-lat]].


\bibitem{Vovchenko:2017xad}
  V.~Vovchenko, A.~Pasztor, Z.~Fodor, S.~D.~Katz and H.~Stoecker,
  Phys.\ Lett.\ B {\bf 775} (2017) 71
  [arXiv:1708.02852 [hep-ph]].


\bibitem{Vovchenko:2017gkg}
  V.~Vovchenko, J.~Steinheimer, O.~Philipsen and H.~Stoecker,
  Phys.\ Rev.\ D {\bf 97} (2018) 114030
  [arXiv:1711.01261 [hep-ph]].


\bibitem{Vovchenko:2018zgt}
  V.~Vovchenko, J.~Steinheimer, O.~Philipsen, A.~Pasztor, Z.~Fodor, S.~D.~Katz and H.~Stoecker,
  Nucl.\ Phys.\ A {\bf 982} (2019) 859
  [arXiv:1807.06472 [hep-lat]].


\bibitem{Lombardo:2006yc}
  M.~P.~Lombardo,
  PoS CPOD {\bf 2006} (2006) 003
  [hep-lat/0612017].


\bibitem{Bornyakov:2016wld}
  V.~G.~Bornyakov, D.~L.~Boyda, V.~A.~Goy, A.~V.~Molochkov, A.~Nakamura, A.~A.~Nikolaev and V.~I.~Zakharov,
  Phys.\ Rev.\ D {\bf 95} (2017) 094506
  [arXiv:1611.04229 [hep-lat]].


\bibitem{Almasi:2018lok}
  G.~A.~Almasi, B.~Friman, K.~Morita, P.~M.~Lo and K.~Redlich,
  arXiv:1805.04441 [hep-ph].



\bibitem{Albright:2014gva}
  M.~Albright, J.~Kapusta and C.~Young,
  Phys.\ Rev.\ C {\bf 90} (2014) 024915
  [arXiv:1404.7540 [nucl-th]].
  
\bibitem{Motornenko:2019arp} 
  A.~Motornenko, J.~Steinheimer, V.~Vovchenko, S.~Schramm and H.~Stoecker,
  arXiv:1905.00866 [hep-ph].
  
\bibitem{CEMtable}
  \href{https://fias.uni-frankfurt.de/~vovchenko/cem_table/}{https://fias.uni-frankfurt.de/\textasciitilde vovchenko/cem\_table/}
  
\end{thebibliography}
\end{document}